\documentclass[10pt]{sig-alternate-10pt}

\usepackage{graphicx}
\usepackage{caption}
\usepackage{subcaption}
\usepackage{pifont}
\usepackage{url}
\usepackage{amsfonts}
\usepackage{times}
\usepackage{threeparttable}
\usepackage{amsmath, amssymb}
\usepackage{pslatex}
\usepackage{enumerate}
\usepackage{clrscode}
\usepackage{arcs}
\usepackage{yhmath}
\usepackage{color}
\usepackage{colortbl}
\usepackage{epstopdf}
\usepackage{array}
\usepackage{flushend}
\usepackage{multirow}
\usepackage{leading}
\usepackage{mathrsfs}
\leading{12pt}
\usepackage{amsmath}
\usepackage[linesnumbered,ruled]{algorithm2e}

\newcommand{\JournalOnly}[1]{}

\begin{document}

\title{SICS: Secure In-Cloud Service Function Chaining}
\numberofauthors{3} 
%
\author{
\alignauthor
Huazhe Wang\\
       \affaddr{University of Kentucky}\\
       \email{huazhe.wang@uky.edu}
\alignauthor
Xin Li\\
       \affaddr{University of Kentucky}\\
       \email{xin.li@uky.edu}
\alignauthor
     Yu Zhao\\
       \affaddr{University of Kentucky}\\
       \email{yzh355@g.uky.edu}
\and
\alignauthor
Ye Yu\\
       \affaddr{University of Kentucky}\\
       \email{ye.yu@uky.edu}
\alignauthor
Hongkun Yang\\
       \affaddr{Google}\\
       \email{yanghk@cs.utexas.edu}
\alignauthor
Chen Qian\\
       \affaddr{UC Santa Cruz}\\
       \email{cqian12@ucsc.edu}
}

\maketitle

\begin{abstract}
There is an increasing trend that enterprises outsource their network functions to the cloud for lower cost and ease of management. However, network function outsourcing brings threats to enterprises' privacy since the cloud is able to access the traffic and rules of in-cloud network functions. Current tools for secure network function outsourcing either incur large performance overhead or do not support real-time updates. In this paper, we present SICS, a secure service function chain outsourcing framework. SICS encrypts each packet header and use a label for in-cloud rule matching, which enables the cloud to perform its functionalities correctly with minimum header information leakage.
Evaluation results show that SICS achieves higher throughput, faster construction and update speed, and lower resource overhead at both enterprise and cloud sides, compared to existing solutions. 


\end{abstract}
\newtheorem{mydef}{theorem}

\section{Introduction}  
\label{sec:intro}
Network functions, also known as middleboxes, are vital parts of modern networks, ranging from security appliances (e.g. firewalls and intrusion detection systems (IDSes)) to performance boosting devices (e.g. HTTP proxies and video transcoder).
Network policies typically require packets to go through a sequence of middleboxes \cite{StEERING,simple}, which is called a \emph{service function chain} \cite{servicechain,StEERING}. For example, all \emph{HTTP} packets should go through $IDS \rightarrow Proxy$; packets from an internal site should be processed by $NAT \rightarrow Firewall$.
Typical enterprise networks employ a large number of middleboxes.
Reported in a survey on middlebox deployment \cite{sherry2012survey}, the total number of various middleboxes approximates that of L2/L3 switches in enterprise networks.
However, dedicated hardware middleboxes come with high infrastructure and management costs, which result from their complex and specialized processing, variations in management tools across devices and vendors.

Stimulated by ever increasing computing power of commodity hardware, the telecommunication industry has proposed network functions virtualization (NFV) \cite{NFVwhite, etsi, att2, opnfv}, which attempts to run software middleboxes atop commodity server hardware.
On the other hand, growing number of applications are hosted in the cloud to reap benefits of it \cite{brocade, juniper, dell}: the statistical multiplexing of hardware resources and hence lower operational cost.
Following this trend, outsourcing middleboxes \cite{APLOMB, gibb2012outsourcing, embark, melisprivate, splitbox} to the third-party cloud now seems promising with the emerging of NFV.
Apart from economical advantages, the flexibility of software and the elasticity of the cloud collectively enable the efficient incorporating of traffic fluctuation, whether stemming from daily pattern or unexpected DDos attacks.
Furthermore, an initial effort \cite{APLOMB}  indicates that outsourcing middleboxes can be achieved without performance impact in terms of latency.

In order to allow the cloud to process traffic, existing middlebox outsourcing solutions \cite{APLOMB, gibb2012outsourcing} require enterprises to provide the cloud with the processing rules for each outsourced middlebox and to redirect their traffic to the cloud in plaintext.
The imposed privacy issues hinder the wide adoption of outsourcing middleboxes to the cloud.
Network traffic data are one of the top business secrets of enterprises and their customers, and most of the enterprise customers are discouraged by the potential data leakage.
Moreover, the processing rules contain sensitive information (i.e., what traffic is not welcome to the enterprise), and its leakage would expose a severe security hole.

Embark \cite{embark} resolves the privacy issue of packet headers and rules of outsourced middleboxes.
To support prefix or range match in some network functions (e.g. firewalls), Embark divides the header space of each field into multiple sub-intervals, which are randomly mapped to other intervals for encryption before the packet enters the cloud.
However, the time to build such sub-interval division requires $O(n^2logn)$, where $n$ is the number of rules.
The situation is aggravated if service chaining is taken into consideration, where the rules in all middleboxes are required to compute the sub-interval division.
Embark does not support real time rule updates, which are very common in practical networks. In the worst case, the encryption of all rules needs to be recalculated.
Moreover, in Embark, field-by-field  encryption of packet headers preserves the order of fields and thus is  vulnerable to brute force attacks.

To address the efficiency and privacy issues of packet header encryption and rule matching, in this paper, we design and implement a middlebox outsourcing scheme SICS, short for secure in-cloud service chaining. SICS protects the private information of packet headers and rules by sending packets with  encrypted headers to the cloud. Since encrypted headers cannot be used for forwarding and rule matching, SICS also assigns a label to each encrypted packet to identify the forwarding and rule-matching behaviors of the packet, which reveals minimal information of in-cloud service chaining. SICS leverages a novel data structure AP Classifier, developed in our previous work \cite{apclassifier},  to support label assignment and dynamic updates, even in the context of service chaining.
\emph{To the best of our knowledge, this middlebox outsourcing scheme is the first system that allows practical in-cloud service function chaining that preserves private information of packet headers and rules and supports fast rule updates.}
Compared to existing solutions, SICS has the following advantages.

\textbf{1. Strong security guarantee.} For each packet, only a label is used to identify its rule-matching behaviors in the cloud. No other information is exposed.


\textbf{2. High-throughput processing at the enterprise side.} Before redirecting packets to the cloud, SICS allows the gateway of the enterprise to find the proper labels with high throughput and little overhead.


\textbf{3. Supporting fast rule update.} The processing rules of middleboxes may be modified frequently \cite{prakash2015pga}.
Moreover, the elasticity of the cloud allows the number of middlebox instances in the cloud to keep on varying to accommodate incoming traffic,
which avoids under-provision during peak load and over-provision for base load \cite{DCNoverload}. SICS supports fast rule updates both at the enterprises  and in the cloud.

\textbf{4. No extra traffic overhead.} The ISP usually charges the traffic redirected to the cloud by volume. SICS does not increase the packet size.

To our knowledge, no prior work can satisfy all above requirements.


Similar to Embark \cite{embark}, SICS focuses on the privacy protection of packet headers and rules. It does not discuss how to perform deep packet inspection (DPI)  to  encrypted payloads.
Existing solutions such as Blindbox \cite{blindbox}  and GPSE \cite{wang2015generalized} are completely orthogonal to our work. SICS can be combined with Blindbox to protect the private information of all packet headers, payloads,  and middlebox rules.  

The rest of the paper is organized as follows. In \S\ref{sec:related}, we present related work. We introduce the system overview in \S\ref{sec:overview}. We present detailed design of the enterprise side in \S\ref{sec:design} and that of the cloud side in \S\ref{sec:middlbeboxes}. We show how SICS supports frequent rule update in \S\ref{sec:update} and analyze its security guarantee in \S\ref{sec:security}. We present the implementation of SICS in \S\ref{sec:Implementation} and evaluation results in \S\ref{sec:evaluation} before concluding in \S\ref{sec:conclusion}.

\section{Related Work}
\label{sec:related}
APLOMB \cite{APLOMB} and Jingling \cite{gibb2012outsourcing} are the pioneer works of  middlebox outsourcing to the cloud.
In particular, extensive experiments were conducted in APLOMB to illustrate the latency inflation due to outsourcing is negligible.
As a parallel work to APLOMB, Jingling  focuses very much on the interfaces and inter-operations between the cloud and customers.
Both works do not take privacy issues into consideration.

Melis et al. \cite{melisprivate} model the behaviors of common middleboxes and proposed a privacy preserving middlebox outsourcing scheme based on fully homomorphic encryption \cite{boneh2005evaluating}.
Though fully homomorphic encryption is generic, the poor performance impedes its use in practice.
Focusing on protecting the privacy of packet headers and middlebox rules, the most closely related work to SICS is Embark \cite{embark}.
Embark uses a method called PrefixMatch to transform packet headers to new IPv6 packet headers.
For each header field, Embark lays out all endpoints of prefixes or ranges on an axis in ascending order.
For all the intervals formed by each consecutive pair of endpoints, Embark calculates the set of prefixes each interval belongs to.
Then Embark assigns an encrypted prefix to each interval. Intervals pertaining to the same set of prefixes receive the same encrypted prefix.
Headers of traffic also employ a per-field encryption. To encrypt a value, Embark first locates the interval and the corresponding IPv6 prefix the value falls in. After that, the value is mapped to a pseudorandom IPv6 address that belongs to the encrypted prefix.
PrefixMatch enables middleboxes running in the cloud to process encrypted packets with encrypted rules.
The main drawback of PrefixMatch is that it does not support real-time updates.
Since network dynamics may occur before completing previous updates, PrefixMatch may never achieve completely correct middlebox processing.
From a security perspective, a field-by-field encryption scheme is vulnerable to brute-force attacks.

In a recent work, Splitbox \cite{splitbox} defines a network function as a pair of a match function and an action function. The action function is distributed to several virtual machines residing in multiple clouds or multiple servers in the same cloud. Computation results from all virtual machines are collected by a local middlebox and final actions of packets are calculated at the local middlebox. The assumption is that an adversary cannot corrupt all virtual machines simultaneously. This assumption does not hold when the attacker is the cloud itself and cloud providers collude with each other.
Match functions are represented as wildcards in Splitbox.
Hash values of match functions and which field of the packet the current match function corresponds to are provided to middleboxes running in the cloud.
Middleboxes calculate hash values of incoming packets and compare them with hash values of match functions to find a match.
When the number of valid bits (bits except wild card bits) of a match function is small, it is easy for adversaries to perform brute-force attacks  to find  pre-images.

\section{System Overview}
\label{sec:overview}

In modern networks, most middleboxes choose proper processing actions (e.g., dropping packets, rewriting packets, or DPI) based on the headers of incoming packets. When a middlebox processes a packet, it finds a rule that  matches the packet header and follows the action of the rule.
Hence,  rule information  specifies packet processing policies of middleboxes. Both  packet headers and rules contain private information of the owner of the enterprise network.
How to perform middlebox outsourcing without exposing the information of packet headers and rules is a challenging problem.
To address this problem, we design and implement SICS, a secure in-cloud service function chaining framework.

\begin{figure*}[t]
\centering
\includegraphics[width=0.8\linewidth]{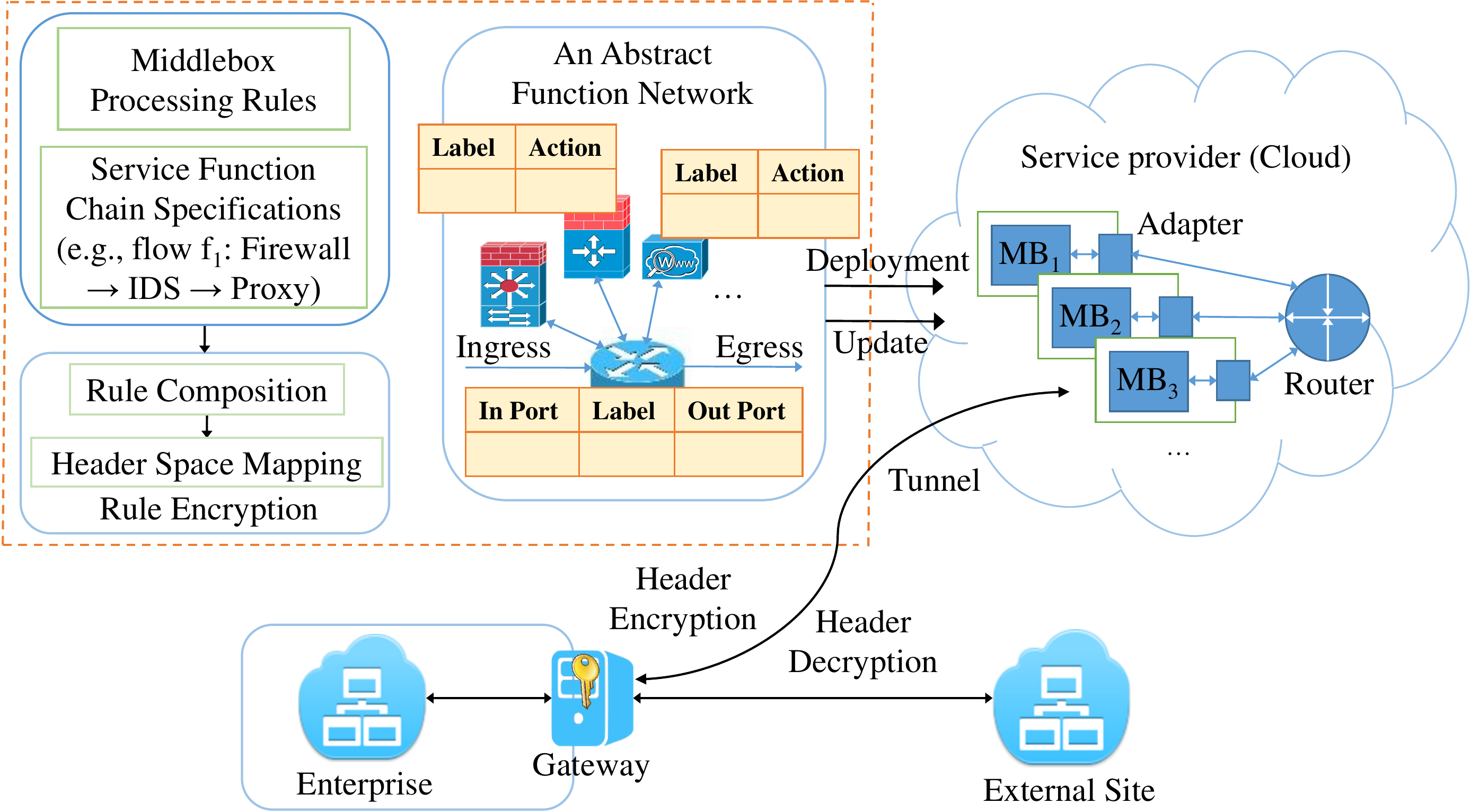}
\caption{Design framework}
\label{fig:framework}
\end{figure*}

\subsection{Outsourcing Architecture}
\label{sec:outsourcing}

We use a network function outsourcing architecture similar to APLOMB \cite{APLOMB}, in which an enterprise redirects its traffic to a third party cloud for network function processing before transmitting them to the internal network or to the Internet. We extend this model to support steering traffic through the desired chains of middleboxes with privacy protection.

The outsourcing architecture contains two parties: an enterprise and a third party cloud providing in-cloud network functions.
A local gateway sitting at the enterprise tunnels both ingress and egress traffic of the enterprise to middleboxes running in the cloud.
To prevent the cloud from accessing headers of the traffic, header fields of incoming packets are encrypted using Advanced Encryption Standard (AES) and sent to the cloud. The in-cloud middleboxes process packets based on encrypted headers and then transmitting them back to the enterprise. Finally, the gateway decrypts the packet and sending them to the internal network.
To enable correct rule-matching with encrypted headers, each packet is assigned with a label. In the cloud, packets are forwarded and processed based on their labels.
When the enterprise communicates with an external site, the operations are similar to those of incoming traffic: outgoing packets from the enterprise are encrypted, sent to the cloud, and sent back to the gateway, before being actually transferred over the Internet.
Note that an optimization that saves on bandwidth and latency can be adopted when communications are between two networks belonging to a same enterprise or two enterprises that has established a secure channel.
After in-cloud processing, the traffic can directly go to the destination site without sending back since the same encryption key is shared by the two networks.

\subsection{Threaten Model}
\label{sec:threatenmodel}

While an enterprise outsources its network functions for decreased cost and ease of management, it brings a challenge on the privacy of header and rule information.
In our threaten model, we assume the cloud to be ``honest but curious'' \cite{introductionsecurity}. The cloud is honest to perform its services correctly. Traffic in cloud is trusted to be forwarded and processed following the requests from the enterprise.
However, the cloud might be curious to learn processing policies at middleboxes or look at the traffic going through.
These data might be sold by disgruntled cloud employees \cite{att, RadioShack} or stolen and exposed by hackers \cite{Chronology, databreach}.
All of these potential threats impede wide adoption of network function outsourcing.
For network function outsourcing, we propose three security properties: (1) The cloud should not learn plaintext of processing rules at in-cloud middleboxes.
(2) For encrypted packets, the cloud should not be able to infer their packet headers based on their in-cloud behaviors. (3) Unlike cryptographic hash functions, label assignment of packet headers does not need to be collision-resistant. In SICS, distinct packets can be assigned with the same label if they have identical behaviors in the cloud.
We assume that the gateway is hosted in the enterprise and trusted. It does not leak information.

\subsection{Middlebox Processing with \\Label Matching}
\label{sec:labelmatching}
In SICS, packet headers are encrypted and hence cannot be used for in-cloud forwarding and rule-matching.
Instead, SICS assigns a label to each packet at the gateway.
Matching fields of the rule tables on in-cloud middleboxes are also specified in labels.
Labels of in-cloud middleboxes are generated and deployed by the network operator at the setup time.
Adopting label-matching in SICS derives from a combination of security and efficiency considerations, as follows:

\textbf{Protect the privacy of headers and rules.}
In our label-matching scheme, packets are forwarded and processed through  in-cloud middleboxes in a sequence specified by service requirements, based on their labels. Forwarding and rule-matching behaviors of packets are completely determined by their labels.
We name all forwarding decisions and behaviors at middleboxes as \emph{network-wide behaviors} of packets.
A set of packets which have the same network-wide behaviors form an \emph{equivalence class}. In SICS, we assign the same label to all packets belonging to the same equivalence class.
Given an encrypted packet with a label, its original packet header cannot be obtained from the label.
For example, there exists a packet header set $S$, packets whose headers are in $S$ share the same network-wide behaviors. A packet is assigned a label ``10110110'' if its header is in $S$. Here, the label is an arbitrary 8 bits-long binary string. The length of a label is
determined by the total number of network-wide behaviors.
Each label only includes two types of information: 1) which middlebox the packet should visit in the cloud, and 2) which action a middlebox should apply to this packet.
%
Rule tables at in-cloud middleboxes are also filled out with label-matching items. If the packet should be processed by an intrusion detection system (IDS), then the rule table of the IDS should contain a rule: ``10110110, IDS''. 
From label-matching tables at middleboxes, the cloud cannot learn original middlebox policies with respect to packet headers.
Note the label-matching method provides minimal information leakage. \emph{The label does not reveal any information other than the packet behaviors of  forwarding and middlebox actions, which are known to the cloud no matter what type of protection is used.}

\textbf{Enable efficient table lookup.}
From a system design perspective, label matching is ideal for service function Chain outsourcing. Labels add little per packet overhead. In our experiments, a 16-bits long label is enough to represent network-wide behaviors in a network with one million 5-tuple rules. The label can be placed at the options field in IPv4 protocol header without incurring extra overhead.
In the rule-matching process, header matching requires to support lookups over hundreds of bits; in contrast, label matching needs only match over some tens of bits. Using a proper designed hash table, label matching can achieve $O(1)$ lookup time without special hardware such as TCAM.
Further, label matching is more flexible than header matching to support network protocol evolution (e.g.,IP addressing from IPv4 to IPv6) and innovations (e.g.,more header fields are involved in matching) which necessitates a change in the matching behavior \cite{MPLSSDN}, especially when the cloud uses hardware forwarding components.

While the use of label matching is not new in a general networking, our specific contributions lie in the design of header space mapping in the context of secure middlebox outsourcing.

\subsection{Design Framework}
\label{sec:framework}

Fig.~\ref{fig:framework} shows the design framework of SICS. Modules in the dashed box are rule preprocessing procedures running on a server of the enterprise.
Inputs are service function chain specifications and processing rules at each middlebox. Service function chains specify middlebox related processing requirements for each kind of traffic. The output is an \emph{abstract function network}.
The abstract function network consists of all middleboxes required in service function chains and a \emph{virtual switch} connecting all middleboxes.
Each middlebox has a rule table which specifies behaviors of packets based on their labels. The virtual switch is equipped with a forwarding table which determines next hops for packets with different labels. Besides label and output port entry, the forwarding table at the virtual switch has an extra entry classifying packets based on their input ports. The input ports are used to identify the segment in the service function chain that the packet is currently in. The result abstract function network should ensure that packets are processed by required middleboxes in a correct sequence. Packets should not pass through middleboxes that are not required.
The enterprise gateway hosts a tunnel between itself and the cloud. It encrypts headers of incoming packets before redirecting them to middleboxes running in the cloud through the tunnel. Headers are restored when packets return from the cloud after service function chain processing.
To address the functionalities outlined above, we design three key modules in SICS:

\textbf{Rule composition:} The rule composition module takes service function chain requirements and middlebox processing rules as input.
It firstly combines certain service requirements to obtain overall middlebox processing sequences for each set of packets.
Based on processing requirements, it generates forwarding rules at the virtual switch to steer packets across middleboxes.
Finally, it uses rule composition algorithms to composite rules at the virtual switch and middleboxes to a list of \emph{predicates}. Each predicate specifies a set of packet headers that has identical behaviors on a certain box \cite{apclassifier}.

\textbf{Header space mapping:} The header space mapping module takes the list of predicates from the rule composition module as input.
It calculates \emph{equivalence classes} using the predicates. Each equivalence class specifies a set of packet headers that has identical behaviors on all boxes.
Hence an equivalence class is the intersection of the packet sets specified by a number of predicates or the negation of predicates.
Each equivalence class is mapped to a  label, which identifies a set of packets belonging to the equivalence class.

\textbf{Packet classification:}
The gateway maintains a packet classifier which classifies packets to equivalence classes based on their headers.
When an equivalence class is found for a given packet, the gateway assigns the label corresponding to the equivalence class to the packet.


\section{Design of Enterprise Gateway}
\label{sec:design}
In this section, we describe design details of SICS enterprise gateway.

\subsection{Rule Composition}
\label{sec:rulecomposition}

A service function chain requires that a certain class of packets should be processed by a number of middleboxes in a designated sequence. For example, all \emph{HTTP} packets should go through IDS $\rightarrow$ Proxy. Packets from an internal site $A$ should be processed by NAT $\rightarrow$ Firewall.
Special considerations should be given on packets which are constrained by more than one service chain requirements.
In previous examples, if an \emph{HTTP} packet comes from the site $A$, it should traverse all middleboxes specified by both chains above.
Service function chains are formulated with respect to a class of packets, specified by packet headers. A set of headers can be represented as a predicate. Variables of the predicate are packet header fields. A predicate $P$ specifies the set of packets for which $P$ is true.
SICS provides a tool to convert packet sets in service function chain requirements to predicates. Consider $m$ processing requirements as the following list:

\[\label{policyenforcement}
   \begin{split}
&S_1, c_1, p_1; \\
&S_2, c_2, p_2; \\
&...\\
&S_m, c_m, p_m.\\
    \end{split}
\]

Let $S_i$ be the predicate specifying the set of packets in the $i$ th requirement. $c_i$ represents the sequence of middlebox processing.
$p_i$ is the priority.
Requirements are listed in descending order of priorities. The priority is necessary to determine the order of middlebox processing when two chains are combined.
In the previous example, the priority is used to determine the overall service function chain for an \emph{HTTP} packet from site $A$ is IDS $\rightarrow$ Proxy $\rightarrow$ NAT $\rightarrow$ Firewall or NAT $\rightarrow$ Firewall $\rightarrow$ IDS $\rightarrow$ Proxy.
To ensure that packets are processed by all required middleboxes, we use Algorithm.~\ref{alg:servicefunctionchain} to calculate overall service function chains for each set of packets specified by a predicate.

\begin{algorithm}
    \SetKwInOut{Input}{Input}
    \SetKwInOut{Output}{Output}

    \Input{Sorted service function chain requirements ($S_i, c_i$ for $i = 1,...,m$).}
    \Output{A list of predicates $\mathscr{F} = \{f_1,f_2,...f_n\}$ and their overall function chains.}
    $Done\leftarrow false$, $\mathscr{F} = \varnothing$,  $\mathscr{T}_1=\varnothing$, $\mathscr{T}_2 = \varnothing$

    \For {$i=1$ to $m$}
        {
            $S_i\leftarrow S_i \wedge \neg Done$, $Done\leftarrow S_i \vee Done$\\
            $\mathscr{T}_1$.add($S_i$), $\mathscr{T}_2 = \varnothing$\\
            \For {$j=i$ to $m$}
            {
                 \For {each $f \in \mathscr{T}_1$}
                 {
                    \eIf{$f \wedge S_j $ = false}
                    {
                        $\mathscr{T}_2$.add($f$)
                    }
                    {
                        $\mathscr{T}_2$.add($f \wedge S_j$)\\
                        append $S_j$'s chain $c_j$ to $(f \wedge S_j)$'s chain\\
                        \If{$f \wedge \neg S_j  \neq false$}
                        {
                            $\mathscr{T}_2$.add($f \wedge \neg S_j$)
                        }
                    }
                 }
            $\mathscr{T}_1 = \mathscr{T}_2, \mathscr{T}_2 = \varnothing$
            }
            $\mathscr{F}.addall(\mathscr{T}_1)$
        }

    \textbf{Return} $\mathscr{F}$

    \caption{Calculate overall service function chains.}
    \label{alg:servicefunctionchain}
\end{algorithm}

The output of Algorithm.~\ref{alg:servicefunctionchain} is a list of predicates $\mathscr{F}$ and service function chains for the packet sets specified by each predicate.
From the algorithm, we see that the conjunction of any two predicates in $\mathscr{F}$ is equal to false (false refers to an empty set), so packet sets specified by any two predicates have no intersection.

\begin{figure}[t]
\centering
\includegraphics[width=1\linewidth]{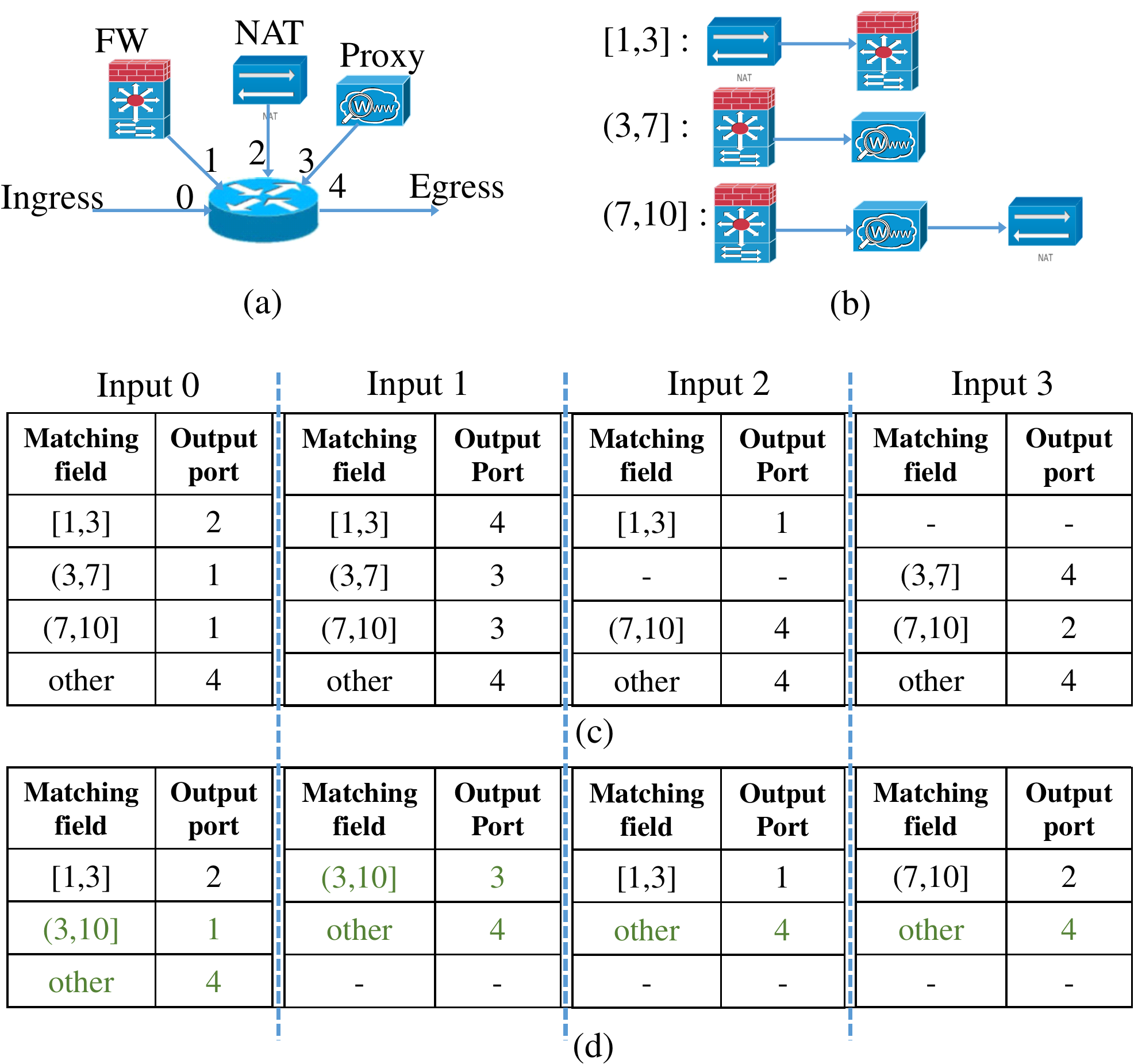}
\caption{(a) An abstract function network. (b) Service function chain requirements. (c) Original forwarding table. (d) Composed forwarding table.}
\label{fig:forwardingtable}
\end{figure}

Next, we fill out the forwarding table of the virtual switch in the abstract function network using the output of Algorithm.~\ref{alg:servicefunctionchain}.
To see how this works, we use an example as shown in Fig.~\ref{fig:forwardingtable}. Fig.~\ref{fig:forwardingtable}(a) is an abstract function network with three middleboxes. All middleboxes are connected by a virtual switch with five ports. Three ports are used to link middleboxes and the other two are ingress and egress port. Fig.~\ref{fig:forwardingtable}(b) shows three sample service function chains. The class of packets in  each chain is specified by an integer range. \footnote{All packet sets are converted to predicates and represented by binary decision diagrams (BDDs) \cite{BDD} in our implementation. Here we use integer ranges for simplicity.}
Fig.~\ref{fig:forwardingtable}(c) is the forwarding table at the virtual switch that steers traffic across middleboxes according to the service chains in Fig.~\ref{fig:forwardingtable}(b). As the forwarding table has an entry input port, we partition the forwarding table into sub-tables, with each corresponds to an input port.
From the Fig.~\ref{fig:forwardingtable}(c), we see that many items in each sub-tables share the same output port.
To reduce the size of each table, ranges which have the same output port are merged.
The result forwarding table is shown in Fig.~\ref{fig:forwardingtable}(d). The total items in the forwarding table are reduced from 14 to 9. When packet sets are represented as predicates, the merge operation can also be efficiently executed by calculating disjunctions of BDDs.
With predicate compositions, there exists at most one predicate for an output port in each sub-table.

Typically, processing rules of middleboxes are generated locally. We use Algorithm.~\ref{alg:middleboxaction} to compose rules of middleboxes before they are deployed in the cloud.
We formulate rules at middleboxes in the form: $(R_i, p_i, a_i)$. $R_i$ denotes the predicate converted from the $i$ th rule. $p_i$ is the priority. $a_i$ specifies the behavior of packets matching this rule. We sort all rules at a middlebox in descending order with respect to priorities.
When a packet is checked against rules at a middlebox, it is matched by the first rule whose predicate evaluates to true for the packet.
We use Algorithm.~\ref{alg:middleboxaction} to compute a single predicate for each behavior at a middlebox.
For a simple example, packets at a firewall only have two possible behaviors: be allowed or denied. Given the access control list (ACL) of the firewall, we compute a single predicate that specifies the packet set allowed by the ACL.

\begin{algorithm}
    \SetKwInOut{Input}{Input}
    \SetKwInOut{Output}{Output}

    \Input{Rules at a middlebox}
    \Output{A list of predicates $\mathscr{P} = \{P_1,P_2,...P_n\}$}

    \For{$i=1$ to $j$}
    {
        $P_j \leftarrow false$
    }
    $Done \leftarrow false$

    \For{$j=1$ to $m$}
    {
        \If{$R_j$ has a same behavior as $P_i$}
        {
           $P_i \leftarrow P_i \vee (R_i\wedge \neg Done)$ \\
            $Done \leftarrow Done \vee R_i$
        }
    }

    \textbf{Return} $\mathscr{P}$

    \caption{Calculate a predicate for each behavior.}
    \label{alg:middleboxaction}
\end{algorithm}

\subsection{Header Space Mapping}
\label{sec:headerspacemapping}

Having obtained a list of predicates for each middlebox as well as the virtual switch in the rule composition module, we use these predicates to calculate equivalence classes within a packet header space.
From previous algorithms, we see that the conjunction of any two predicates from the same box is false while disjunction of all predicates is true. (We have added a default predicate at each box.) Predicates from a box can be seen as a partition which divides the packet header space into several sub-spaces. For a middlebox, a sub-space specifies a set of packets having a same behavior at the middlebox.
For the virtual switch, a sub-space presents a packet set with a same forwarding output port.
If we place predicates from all boxes together, the partition of the header space will become combinatorially finer due to intersections of predicates from different boxes.

\begin{figure}[t]
\centering
\includegraphics[width=1\linewidth]{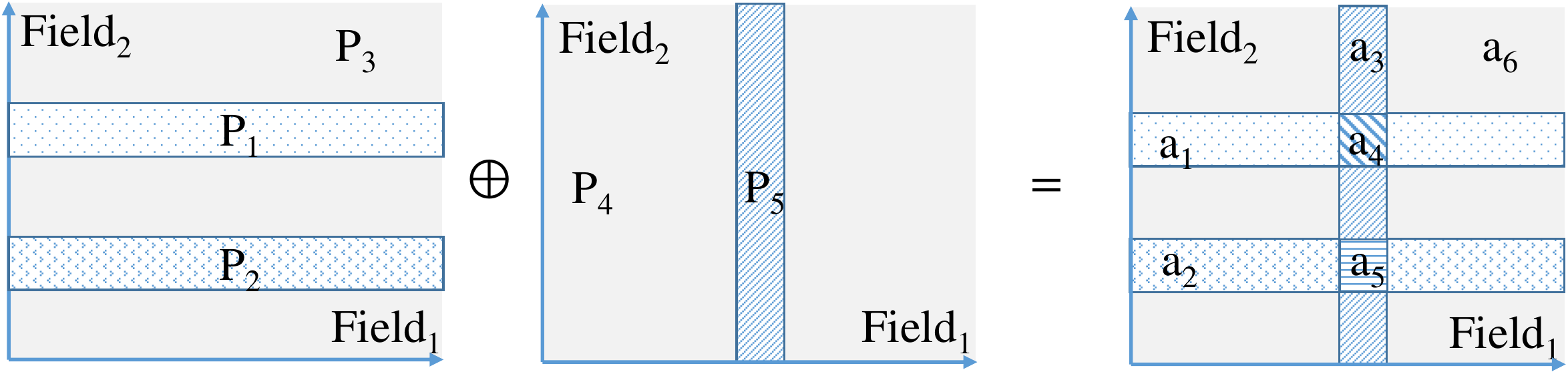}
\caption{Header space divided by predicates}
\label{fig:headerspace}
\vspace{-1ex}
\end{figure}

Fig.~\ref{fig:headerspace} shows an example illustrating the process of combining predicates from two boxes.
In this example, each predicate is associated with two header fields. Five predicates $P_1 \sim P_5$ from two boxes are placed together in one packet header space. Then the header space is partitioned into fifteen blocks.
Each block represents a set of headers belonging to an identical set of predicates. That is, packet headers within one block will match same predicates and then have identical behaviors at all boxes.
In \S\ref{sec:labelmatching}, we define network-wide behaviors of a packet as its overall forwarding paths and behaviors at middleboxes in the network.
So packets within one block of the header space have same network-wide behaviors. Recalling the definition of the equivalence class in \S\ref{sec:labelmatching}, we can conclude that headers within one block belong to the same equivalence class.
Note that an equivalence class is not necessarily continuous. As shown in Fig.~\ref{fig:headerspace}, the partition of fifteen blocks has six equivalence classes $a_1 \sim a_6$.

Given a list of predicates, we calculate its equivalence classes using algorithms in \cite{APVerifier}.
The set of equivalence classes has two key properties: (1) Packets within an equivalence class have identical network-wide behaviors. That is, these packets will traverse the same sequence of middleboxes and have same behaviors at each middlebox in the network.
(2) Each input predicate is equal to disjunctions of a subset of equivalence classes. As shown in Fig.~\ref{fig:headerspace}, $P_1 = a_1 \vee a_4$ and $P_5 = a_3 \vee a_4 \vee a_5$.

With these properties, it is intuitive to map the packet header space with respect to equivalence classes.
We map packet headers within an equivalence class to one label. The label in our implementation is a binary string. The length of the string is determined by the total number of equivalence classes. A 16-bits long label can support a network with up to 65536 equivalence classes, with each corresponds to a unique network-wide behavior. In rule tables of in-cloud boxes, a predicate $P$ is represented as a set of labels. The labels are determined by the subset of equivalence classes whose disjunction is $P$.

\subsection{Packet Classification}
\label{sec:gateway}

Before the gateway assigning labels to packets, we need to figure out which equivalence class a given packet belongs to.
Equivalence classes are specified by predicates. A straightforward approach is to test the packet header against these predicates linearly.
However, this approach is slow. In our experiments on the Stanford network \cite{Stanford}, which has about 500 equivalence classes, only 20K queries per second was achieved.
To make SICS practical in service function chain outsourcing, a challenge is to design a data structure that could classify packets to their equivalence classes with a high throughput. The data structure should not incur large memory overhead at the enterprise. More importantly, it should support real time update to deal with dynamics resulting from middlebox load balancing or changing of processing policies.

In SICS, we leverage algorithms in \cite{apclassifier} and all predicates obtained from the rule composition module to construct a packet classifier.
In the query process, the output of the packet classifier is the equivalence class the given packet belongs to.
With the rule composition, the total number of input predicates is much smaller than the number of rules from middleboxes and the virtual switch, which reduces the time cost to construct the packet classifier.
In our experiments, the construction time of the packet classifier for a ruleset with 100K rules is less than one second.
With the optimization technique in \cite{apclassifier}, the packet classifier can find an equivalence class for a given packet by testing the packet header against only tens of predicates.
Further, as shown in \cite{apclassifier}, the packet classifier can be updated in real-time with very little overhead.


\section{Design of In-cloud Middleboxes}
\label{sec:middlbeboxes}
The goal of SICS is to protect the privacy of packet headers and middleboxes processing rules.  SICS can be used to module middleboxes which determine packet behaviors based on packet headers.
We call these middleboxes as header-related middleboxes (e.g., firewalls, NAT, and L4 load balancing).
For middleboxes which also check other packet fields, such as payload (e.g., proxies and IDSes), SICS can be combined with recent works of secure DPI \cite{blindbox, wang2015generalized}.

To encrypt processing rules of middleboxes, the rule composition module has converted rules at each middlebox to a list of predicates.
With the property of equivalence classes, each predicate is represented as a set of labels.
At the enterprise gateway, a packet is assigned a label corresponding to the equivalence class the packet belongs to.
So an in-cloud middlebox can determine the behavior of a packet by searching its label against the label set of each predicate to find a match.
Next, we discuss header-related middleboxes in two categories: static middleboxes that do not modify packet headers and others work as header transformers which rewrite some or all fields of packet headers.

\textbf{Static middleboxes.}
Static middleboxes process packets without modifying headers.
We can use one label for a packet to identify its processing at all static middleboxes it traverses.
At a middlebox, labels determine packet behaviors. They are saved in the form of $(Key, Value)$ pairs in a hash table. A key represents a label and its value store the behavior of packets with that label.
If a middlebox finds a match for the label of a packet, the behavior of the packet at the middlebox is obtained.

For stateful middleboxes that record information about connections, encrypted packet headers are used to distinguish flows with the same label.

\textbf{Dealing with header transformers.}
When an in-cloud middlebox modifies packet headers, behaviors of packets on downstream boxes must be determined by their new headers.
In label-matching, the rest packet behaviors should be instructed by new labels.
For example, when an internal packet goes through a NAT, its source IP is modified to the external IP of the NAT (The source port is changed randomly. The randomly chosen source port does not influence packet behaviors, so it is not considered in assigning labels).
The NAT needs to assign a new label to the packet since the header of the packet has been changed.
However, it is hard for in-cloud middleboxes to calculate new labels for packets they have just modified.
This is because labels are assigned to packets at the gateway based on all their header fields.
To keep the privacy, headers of packets are encrypted before they enter the cloud. It is impossible for in-cloud middleboxes to calculate labels for packets without knowing their headers.
Also, new modified header fields should be encrypted and not be accessed by the cloud.
So middleboxes which modify packet headers need to assign new labels to packets without learning both their original and new headers.

\begin{figure}[t]
\centering
\includegraphics[width=1\linewidth]{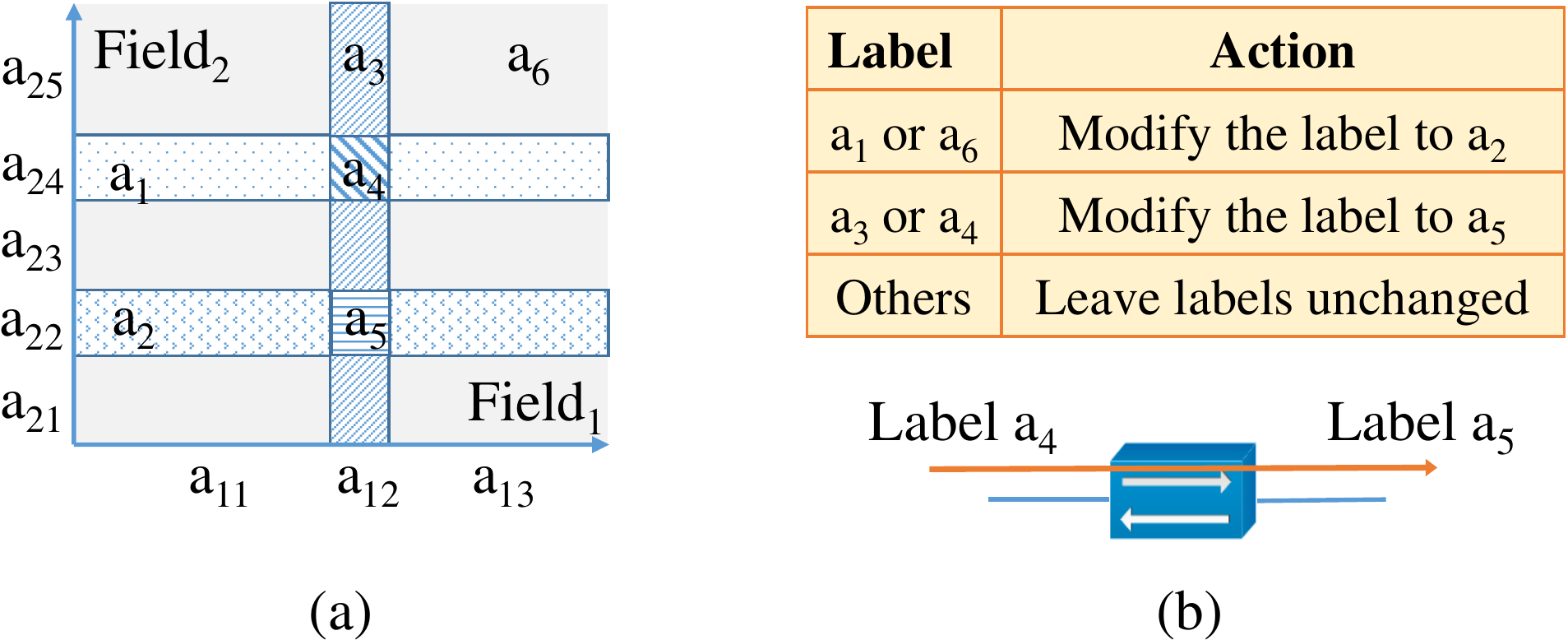}
\caption{(a) Per-field equivalence class. (b) An example on header modification.}
\label{fig:headermodification}
\vspace{-2ex}
\end{figure}

To address the problem above, we design a label-to-label replacement strategy. In \S\ref{sec:headerspacemapping}, we partition the header space as a whole to equivalence classes. A packet is classified to an equivalence class based on all its header fields.
Nonetheless, there also exists a partition along each field of the header space. We call one part of the partition as a per-field equivalence class.
To illustrate this concept, we revisit the previous example in Fig.~\ref{fig:headermodification} (a). The header space has six equivalence classes, $a_1 \sim a_6$. Along each field, the header space is also partitioned into per-field equivalence classes.
Along $field_1$, there are $a_{11} \sim a_{13}$. Along $field_2$, there are $a_{21} \sim a_{25}$.
A property is that an equivalence class can be identified by a set of per-field equivalence classes, with each took from one field.
In Fig.~\ref{fig:headermodification} (a), $a_4$ can be identified by $a_{12}$ and $a_{24}$ and $a_5$ can be identified by $a_{12}$ and $a_{22}$.

Based on the property of the per-field equivalence class, a label replacement table can be produced for each equivalence class when one header field is modified.
As shown in Fig.~\ref{fig:headermodification} (b),
we assume that the function of the middlebox is rewriting $field_2$ of all incoming packets to a new value $v$ and the value $v$ belongs to the per-field equivalence class $a_{22}$.
We can get a label replacement table based on logical relationships between equivalence classes as shown in Fig.~\ref{fig:headermodification} (b).
For a packet with a label $a_4$, the orange line shows its behavior when it passes the middlebox.
Its label is modified to $a_5$ since it matches the second line of the table. The rest behaviors of the packet will be guided by the label $a_5$.
Note that all contents of the label replacement table are generated by the gateway and sent to the cloud at setup time.

Besides modifying labels, the middlebox assigns an index from a header field pool to the packet. It receives these indexes from the gateway at setup time. Each index is reserved for a rewritten header field. When the gateway receives a packet with such an index, it restores the rewritten header fields.

\section{Update operations}
\label{sec:update}


In-cloud middleboxes may experience bandwidth fluctuations \cite{opennf} and middlebox overload is a common cause of failures \cite{DCNoverload, APLOMB}. Thus, it necessary to balance the load across middleboxes in real time.
Moreover, service function chain requirements and processing rules of middleboxes may need to be modified over time.
All of these dynamics results in rule updates at the virtual switch and middleboxes.
To keep the correctness and performance of the in-cloud processing, it is necessary for a service function chain outsourcing framework to support real-time update. We use the method presented in \cite{APVerifier} to convert a rule insertion or deletion to predicate changes. If there exist predicate changes after a rule update, we perform the method presented below to update both the gateway and in-cloud boxes.

\subsection{Update at the Gateway}
\label{sec:updategateway}

When a predicate is added or deleted, the update procedure starts from updating the packet classifier at the gateway. We use algorithms from \cite{apclassifier} to execute the update.
If equivalence classes are changed by an update, the packet classifier will start to classify packets to the new set of equivalence classes after the update.
Updates of the packet classifier can be executed very fast. In our experiments, the average cost of adding a predicate is less than 0.5 ms.

To figure out the influence of an update to existing predicates, the gateway maintains a representation list for each predicate. The list of a predicate includes all equivalence classes whose disjunction is equal to the predicate.
In the previous example, the representation list of $P_5$ is $\{a_3, a_4, a_5\}$ and that for $P_2$ is $\{a_2, a_5\}$.
The table is adjusted dynamically based on the current state of each predicate. An update may result in modifications of multiple representation lists.
When the representation list of a predicate is modified, the gateway sends update instructions to the source box of the predicate running in the cloud.

\subsection{Update in the Cloud}
\label{sec:updatecloud}

Predicate updates in the cloud are treated in four cases. For the cases in which a new predicate $P'$ is added, we assume that an existing equivalence class $a_i$ is partitioned into two equivalence classes $a_i$ and ${a_i}'$.

Case 1: Adding a new predicate $P'$ at a middlebox. We firstly calculate the representation list of $P'$.
For each equivalence class in the representation list, we insert its label into the hash table of the middlebox. Values inserted are the behavior identified by $P'$.
For any predicate $P$ whose representation list originally contains $a_i$, we add a new equivalence class ${a_i}'$ into its list.
Meanwhile, the label corresponding to ${a_i}'$ is inserted into the hash table of $P$'s source box in the cloud, the value is $P$'s behavior.

Case 2: Deleting an existing predicate $P$ at a middlebox. Labels corresponding to equivalence classes in the representation list of $P$ are deleted from the hash table of the middlebox.
No other operations are needed.

Case 3: Adding a new predicate $P'$ at the virtual switch. Similar as in case 1, labels corresponding to $P'$ and modifications for other predicates are inserted.
Adding a new predicate at the virtual switch happens in two scenarios: a) A new middlebox processing is inserted into the service function chain, packets are forwarded to the new middlebox before they are processed by the rest of function chain;
b) To balance the load, instead of being processed by the specified middlebox, a portion of traffic is rerouted to a new middlebox of the same function.
In case (a), the new added predicate $P'$ is marked as ``new''. Packets matching $P'$ will not change input ports and continue their service chain after being processed by the middlebox identified by $P'$. In case (b), the new added predicate $P'$ is treated as a normal predicate.

Case 4: Deleting an existing predicate $P$ at the virtual switch. Similar as in case 2, labels corresponding to $P$ are deleted. The deleted predicate $P$ is marked as ``deleted''. Incoming ports of packets matching $P$ are changed to the port determined by $P$ instantly without actually being forwarded to the middlebox.

Note that, for stateful middleboxes, states recorded will not be disturbed in updating since we distinguish flows using their encrypted header fields.

\subsection{Maintain the Consistency}


Rule updates need to be treated carefully. Inconsistent state between the gateway and boxes in the cloud may lead to incorrect middlebox processing.

To maintain per-packet consistency, we buffer packets at the gateway when there come updates. The gateway installs updates as in \S\ref{sec:updategateway}. Then the gateway calculates update schemes for boxes involved and sends update messages to the cloud.
Upon receiving update instructions, boxes in the cloud finish processing all ongoing packets in the pipeline before installing updates.
After that, the gateway begins to process new packets.

To maintain flow consistency, we employ the migration avoidance mechanism in \cite{e2}. New flows are steered to new middlebox instances while existing flows still processed by old ones.
Additionally, a lookup table recording previous flows and their labels are maintained at the gateway. Incoming packets search the lookup table before querying the gateway. If a match occurs, a packet belonging to a current connection will always be assigned the same label.
An item in the lookup table is removed if the connection is terminated.

\section{Security Guarantee}
\label{sec:security}

In SICS, prefixes and ranges from middlebox processing rules are converted to a list of predicates. Then each predicate is represented as a set of labels. Labels are used as matching fields to enable in-cloud functionalities. Labels do not leak size, order or border of predicates.
Also, the cloud cannot learn which field of the packet header a matching corresponding to. 
SICS encrypts packet headers and assigns a label to each packet to identify its in-cloud processing. 
Given an encrypted packet with a label, its original packet header cannot be reversed from the label.
For any two packets that are assigned the same label, the cloud only learns that the two packets have same in-cloud behaviors, but it cannot learn any other information about their orders or values.

Next, we compare the security of SICS with the PrefixMatch in Embark \cite{embark} under a brute force attack. We assume that an attacker (e.g., cloud itself or a hacker) selectively sends sample packets to the gateway and observes their in-cloud behaviors, trying to figure out the plaintext of rules at a  middlebox.
PrefixMatch adopts a per-field encryption scheme. Prefixes or ranges for each header field are encrypted separately.
For an encrypted prefix or range, the attacker knows which field of the packet header the prefix or range corresponding to.
He can obtain the plaintext of the encrypted prefix or range by traversing the entire search space of that field.

An example attack is as following: for the destination port field in IPv4 header, PrefixMatch encrypts a port number interval $[s,e]$ to a random interval $[S,E]$. All port numbers falling in $[s,e]$ are encrypted to values in $[S,E]$. Knowing the interval $[S,E]$, it takes an attacker at most $2^{16}$ queries (e.g., sample packets with destination port traversing from $0$ to $2^{16}$) to find all port numbers in $[s,e]$, where 16 is the length of the port field.
So the attacker has successfully deciphered the encrypted interval $[S,E]$ in the cloud.
In addition, when a future packet matches the interval $[S,E]$, the attacker learns that the original destination port of the packet falls in $[s,e]$.
Similarly, the attacker could get mapping relationships for other fields. As a chosen packet header can test each header field simultaneously,
the number of required queries to decipher all header fields is determined by the length of the longest header field. For five tuples, the longest header field is 32 bits. 
So it takes at most $2^{32}$ quires to decipher a five-tuple based ruleset encrypted using PrefixMatch.

As described in \S\ref{sec:headerspacemapping}, SICS encrypts packet header fields as a whole. That is, all packet header fields are involved in the header space mapping process, the label of a packet is determined by all bits of its header. To launch the same attack described above,
it costs $2^{104}$ queries which is much larger than $2^{32}$ .

\section{Implementation}
\label{sec:Implementation}

\begin{figure}[t]
\centering
\includegraphics[width=\linewidth]{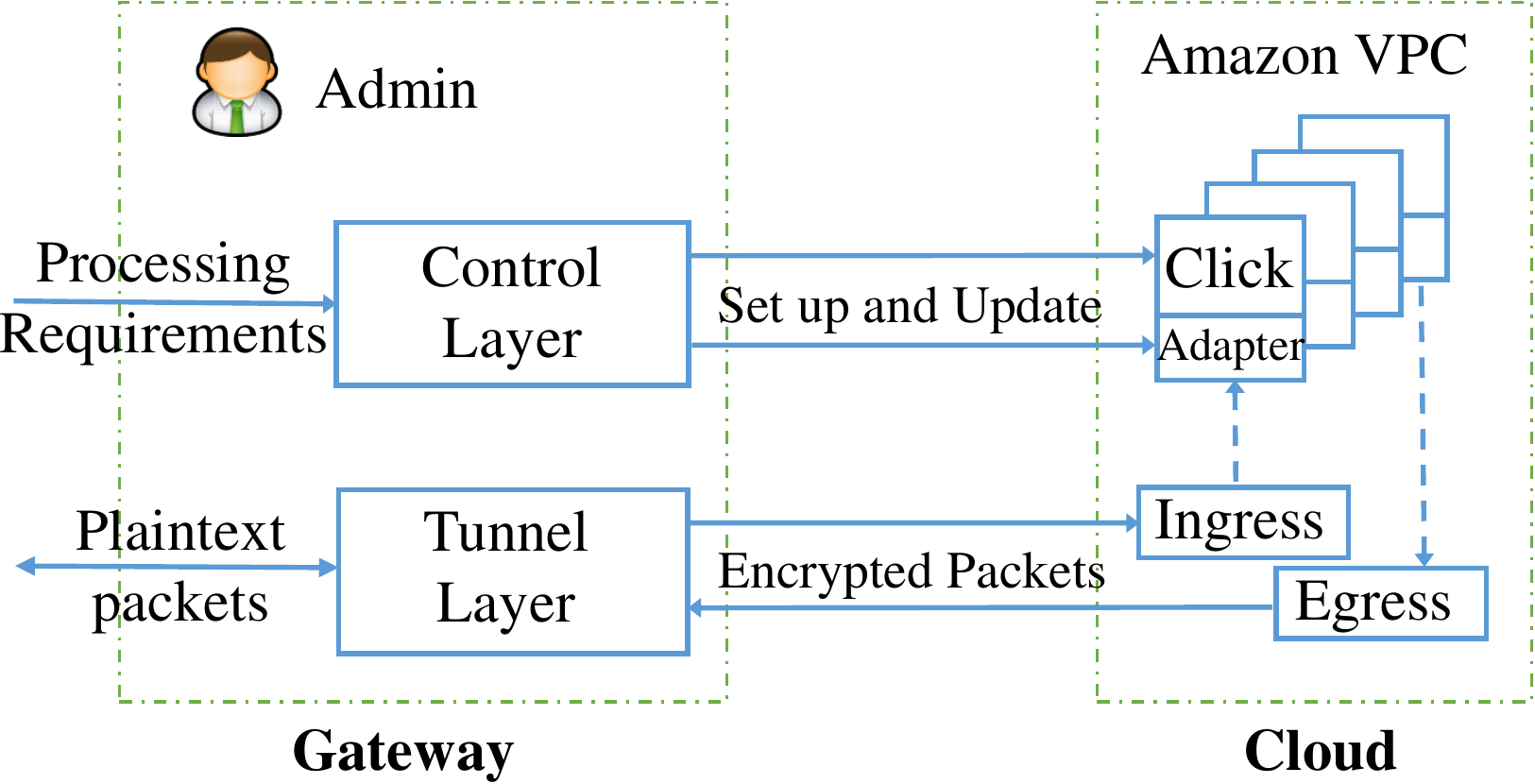}
\caption{Software architecure}
\label{fig:implementation}
\vspace{-3ex}
\end{figure}

We built a prototype of SICS using middleboxes running in Amazon Virtual Private Cloud (VPC) \cite{VPC} and a gateway running on a general purpose desktop computer in the authors' lab. We only use off-the-shelf components provided by existing cloud and end hosts. This makes our design easy to deploy and use.

Fig.~\ref{fig:implementation} shows the software architecture of SICS.
The gateway sitting at the enterprise consists of two layers: a control layer and a tunnel layer.
The control layer takes service function chain requirements and processing rules of middleboxes as input to calculate an abstract function network. The abstract function network indicates the total number of required middleboxes and set up configurations of each middlebox.
The control layer also maintains a packet classifier which classifies packets from the tunnel layer to different labels.
When dynamics happen, the control layer updates the packet classifier in real time. Simultaneously, it calculates necessary updates in the cloud.
We wrote scripts at the control layer to send batched instructions of both setting up and updates to middleboxes running in the cloud.

The tunnel layer hosts packet manipulation, header encryption and VPN tunnels connecting remote instances in the cloud.
The tunnel layer receives plaintext packets from customer networks. It adds labels assigned by the control layer on packets. Our current implementation adds a 16-bits long label into the options field of an IPv4 header. Then, header fields of packets are encysted using symmetric encryption (e.g., AES) . For tunnels, we use OpenVPN \cite{VPN}, a widely-deployed VPN solution working under most operation systems. Encrypted packets are sent out through tunnels using static routes.

\begin{figure}[t]
\centering
\includegraphics[width=\linewidth]{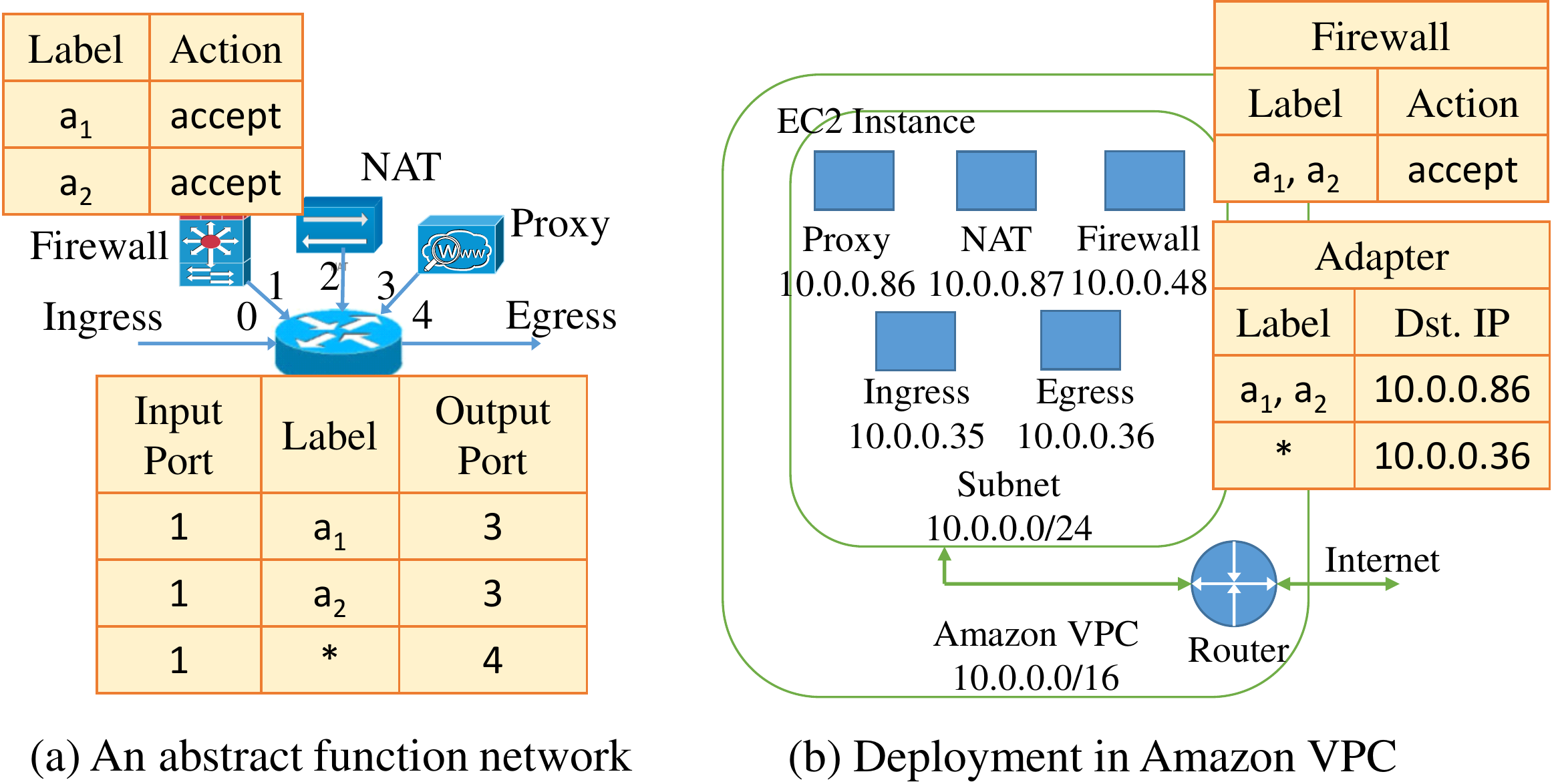}
\caption{Example to convert an abstract function network to in-cloud deployment}
\label{fig:deployexample}
\vspace{-2ex}
\end{figure}

At the cloud side, the abstract function network obtained by the control layer of the gateway can be easily converted into practical deployment in Amazon VPC. Amazon VPC provides a virtual network of EC2 instances. Packets can be transmitted between instances in a VPC using Layer-3 forwarding.
Our deployment in the VPC has an ingress instance, an egress instance,  and other instances, each corresponds to a middlebox. The ingress/egress instance hosts a tunnel endpoint and forwards packets from and to middleboxes running in the cloud.
To enable in-cloud packet forwarding, we add an adapter layer at each EC2 instance to encapsulates and decapsulate incoming and outgoing packets.
The adapter layer sits between ethernet devices and Linux kernel network stack which minimizes modifications to current middleboxes.
Before a packet is processed by a middlebox, the adapter layer removes its ethernet and outermost IP header.
After the middlebox processing, the adapter layer encapsulates the packet with new headers which guides the packet to its next required middlebox processing based on its label.
With the abstract function network, the adapter layer learns the next required middlebox processing of a packet by searching its label against the forwarding table of the virtual switch in the abstract function network.
As shown in Fig.~\ref{fig:forwardingtable}, the forwarding table is partitioned into sub-tables with respect to input ports. For the adapter layer at a middlebox, it can get new headers of packets by only searching for the sub-table corresponding to the middlebox.
Fig.~\ref{fig:deployexample} presents an example of converting an abstract function network into practical in-cloud deployment.
Fig.~\ref{fig:deployexample} (a) is an example abstract function network with three middleboxes. For simplicity, only the rule table and the forwarding sub-table corresponding to a firewall are given.
As shown in Fig.~\ref{fig:deployexample} (b), the abstract function network is converted to a VPC subnet consisting of five EC2 instances.
For the firewall, its rule table is obtained directly from the one in the abstract function network. The adapter lookup table at the firewall derives from the forwarding sub-table which has its input port equals to 1.
From this table, we see that packets with label $a_1, a_2$ should be forwarded to port 3 which corresponds to a proxy. Other packets are forwarded to port 4 which is the egress. In the practical deployment, packets with label $a_1, a_2$ are encapsulated with the proxy's private IP 10.0.0.86 as their destination IP addresses. Other packets are encapsulated with IP 10.0.0.36 which corresponds to the egress.

For middlebox processing, we use Click \cite{click}, a customized modular router. We build Click elements to perform a firewall and a NAT.
The only part that needs to be modified is the matching process of middleboxes' rule tables.
We have implemented a label-matching element using Cuckoo hash table \cite{cuckoo,cuckoofilter}. The Cuckoo hash table stores $(Key, Value)$ pairs with a key representing a label and a value indicating behaviors of packets with that label.

\section{Experimental Evaluation}
\label{sec:evaluation}
We evaluate the performance of SICS. We first present the performance of the gateway (\S\ref{sec:gateway}). Then we analyze bandwidth overhead due to the encryption scheme in SICS (\S\ref{sec:bandwidth}). Finally, we demonstrate SICS's in-cloud performance, dynamic scaling capability and its resilience to middlebox failures (\S\ref{sec:middlebox}).
The gateway is built on a general purpose desktop computer with quadore@3.2G and 16GB memory. It redirects traffic from another machine with the same model. For most experiments, we use a synthetic workload generated by the Pktgen \cite{Pktgen}.
We use data plane states from two real networks \cite{Internet2} and \cite{Stanford} as starting
points to create service function chain requirements and rulesets at middleboxes.
In our experiments, all policies are based on five tuples (Source IP, Source Port, Destination IP, Destination Port, Protocol).

\subsection{Gateway}
\label{sec:gateway}

We first evaluate the performance of the gateway. We compare with PrefixMatch in Embark \cite{embark} implemented by us.

\begin{figure}[t]
\centering
\includegraphics[width=0.8\linewidth]{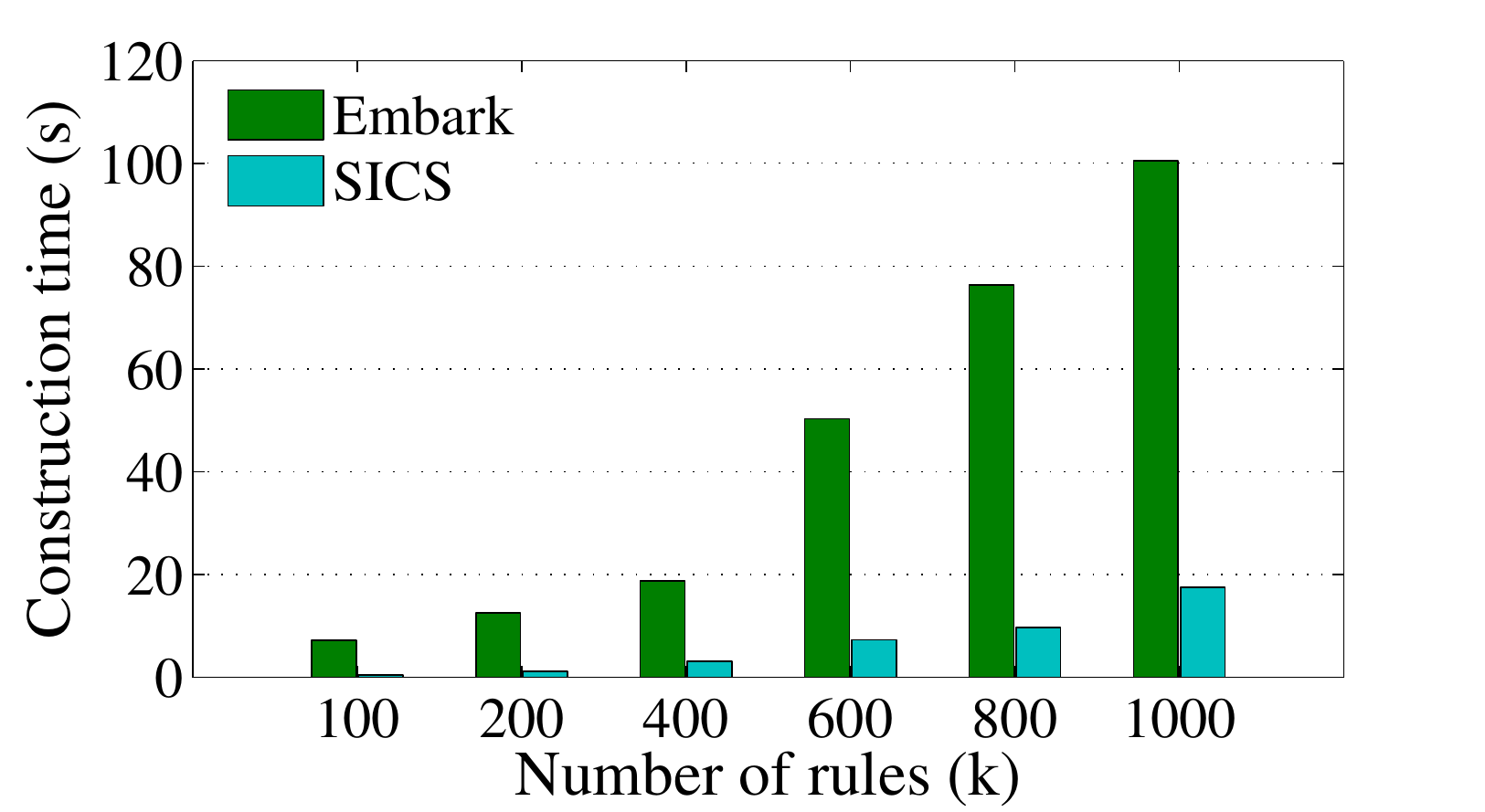}
\caption{Construction time as \# of rules increases.}
\label{fig:construction_time}
\vspace{-1ex}
\end{figure}

\textbf{Construction time.} Fig.~\ref{fig:construction_time} shows construction time of the gateway with the size of ruleset increasing from 100K to 1000K. The construction time in SICS includes time cost of rule composition, computing equivalence classes and construction of the packet classifier. In Embark, it is the time cost to construct the data structure for PrefixMatch.
The PrefixMatch in Embark works only on one header field, so PrefixMatch needs to be run for each header field one after another. In Fig.~\ref{fig:construction_time}, we see that the time cost of PrefixMatch in Embark is at least 5 times larger than SICS for all six rule set sizes.
The reason lies in the facts that the total number of sub-intervals for each header field in PrefixMactch is much larger than the number of equivalence classes in SICS. In our experiments, the Internet2 data set which has approximately 70K rules produces 216 equivalence classes. The number of sub-intervals calculated using PrefixMatch is more 9000.
This makes the process to find intervals pertaining to the same set of prefixes in PrefixMatch very inefficient, especially when the size of the ruleset is large.
As shown in Fig.~\ref{fig:construction_time}, constructing of the gateway in SICS only uses 383.7 ms for the ruleset with 100K rules. It is still less than 20 s when the size of the ruleset increases to 1000K.

\begin{table}[htbp]
\caption{Construction time of the gateway.}
\centering
\small
\begin{tabular}{|c|c|c|c|}
\hline
\begin{tabular}[c]{@{}c@{}}Size of \\ the Ruleset \\ (K)\end{tabular} & \begin{tabular}[c]{@{}c@{}}Rule \\ Composition \\ (ms)\end{tabular} & \begin{tabular}[c]{@{}c@{}}Computing \\ Equivalence \\ Classes (ms)\end{tabular} & \begin{tabular}[c]{@{}c@{}}Constructing \\ Packet \\ Classifier (ms)\end{tabular} \\ \hline
100  & 315.4                                                               & 14.9                                                                            & 53.4                                                                            \\ \hline
200  & 1118.3                                                              & 15.2                                                                            & 83.2                                                                            \\ \hline
400 & 2984.5                                                              & 22.4                                                                            & 129.0                                                                           \\ \hline
600 & 7125                                                                & 25.2                                                                            & 148.2                                                                           \\ \hline
800 & 9474.3                                                              & 30.5                                                                            & 249.8                                                                           \\ \hline
1000 & 17168.4                                                             & 41.3                                                                            & 313.9                                                                           \\ \hline
\end{tabular}

\label{table:construction}
\end{table}

Next, we investigate time cost of each phase in gateway construction separately.
As shown in Table.~\ref{table:construction}, rule composition accounts for most of the overhead while computing equivalence classes and classifier construction can be finished in tens of millisecond.
When the size of the ruleset increases, rule composition costs more time since more rules are involved in the computation.
However, the number of predicates and equivalence classes do not increase much since there usually exists large amounts of redundancy as well as similarity between rules at middleboxes \cite{APVerifier-TR}. So the time used to compute equivalence classes and construct the packet classifier do not increase significantly.
Note that the gateway in SICS needs to be constructed for only once at setup time. After that, the packet classifier can be updated incrementally when there exist rule changes.

\begin{figure}[t]
\centering
\includegraphics[width=0.8\linewidth]{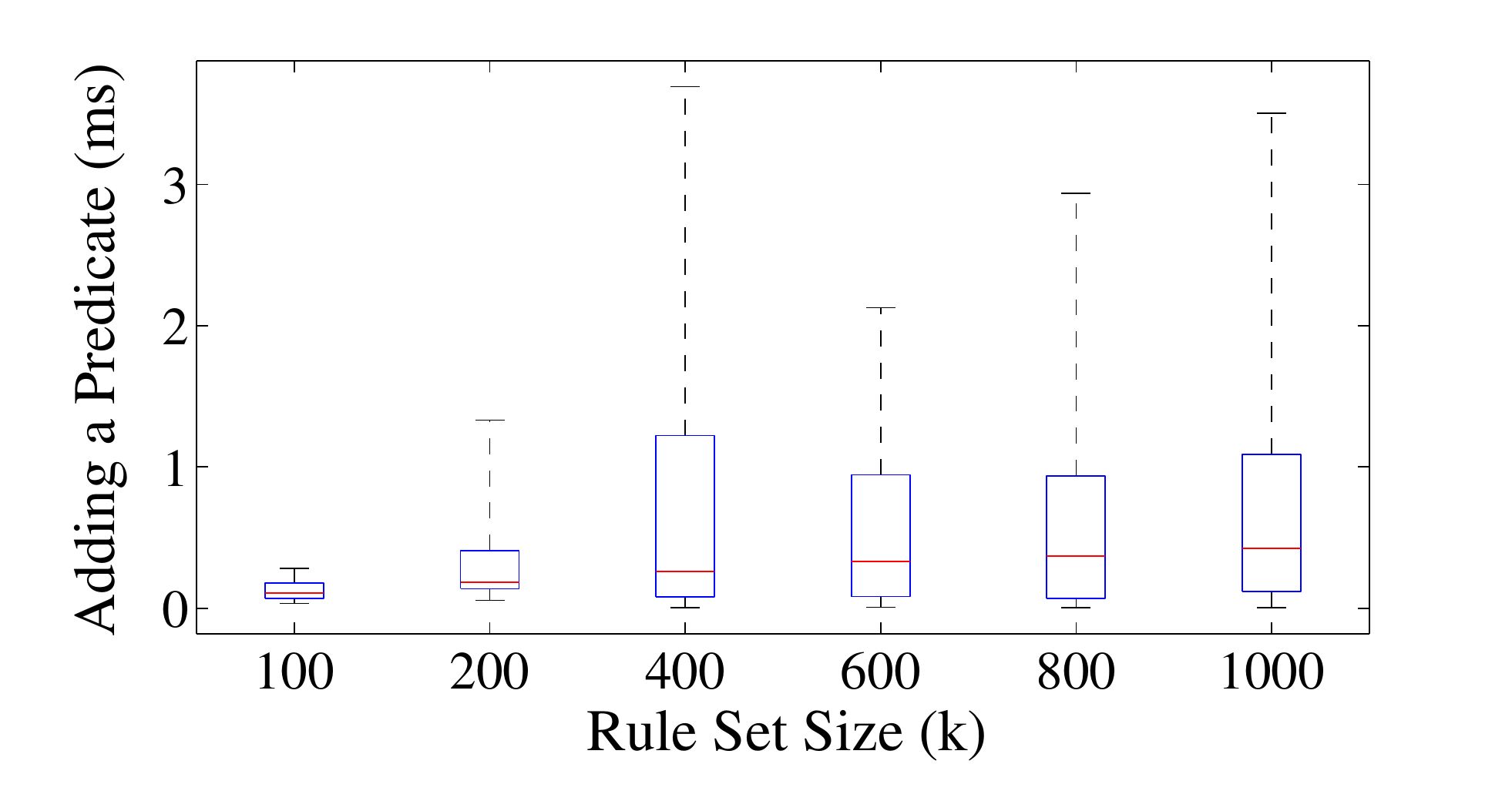}
\caption{Box plot of update cost for each rule set.}
\label{fig:update}
\vspace{-1ex}
\end{figure}

\textbf{Real-time update of the packet classifier.}
In this set of experiments, we first construct the packet classifier using a subset of predicates and then keep adding new predicates. We measure the time cost to add each  predicate and update the packet classifier.
Fig.~\ref{fig:update} shows the distribution of time cost for adding a predicate for the ruleset with the number of rules increasing from 100K to 1000K. From the figure, we see that the medium cost for adding a predicate does not have oblivious differences when the size of ruleset increases.
It is less than 0.5 ms for all rulesets. The worst case may take 2-2.5 ms when the size of the ruleset is large.
Deleting a predicate does not require extra computation, hence there is no result for deletions.

PrefixMatch in Embark may need to be reconstructed when a rule update happens. So the time cost of a rule update is the same as its  construction cost, which can be approximately 100 s.
Since network dynamics may occur during the reconstruction, PrefixMatch may never achieve completely correct middlebox processing.

\begin{figure}[t]
\centering
\includegraphics[width=0.8\linewidth]{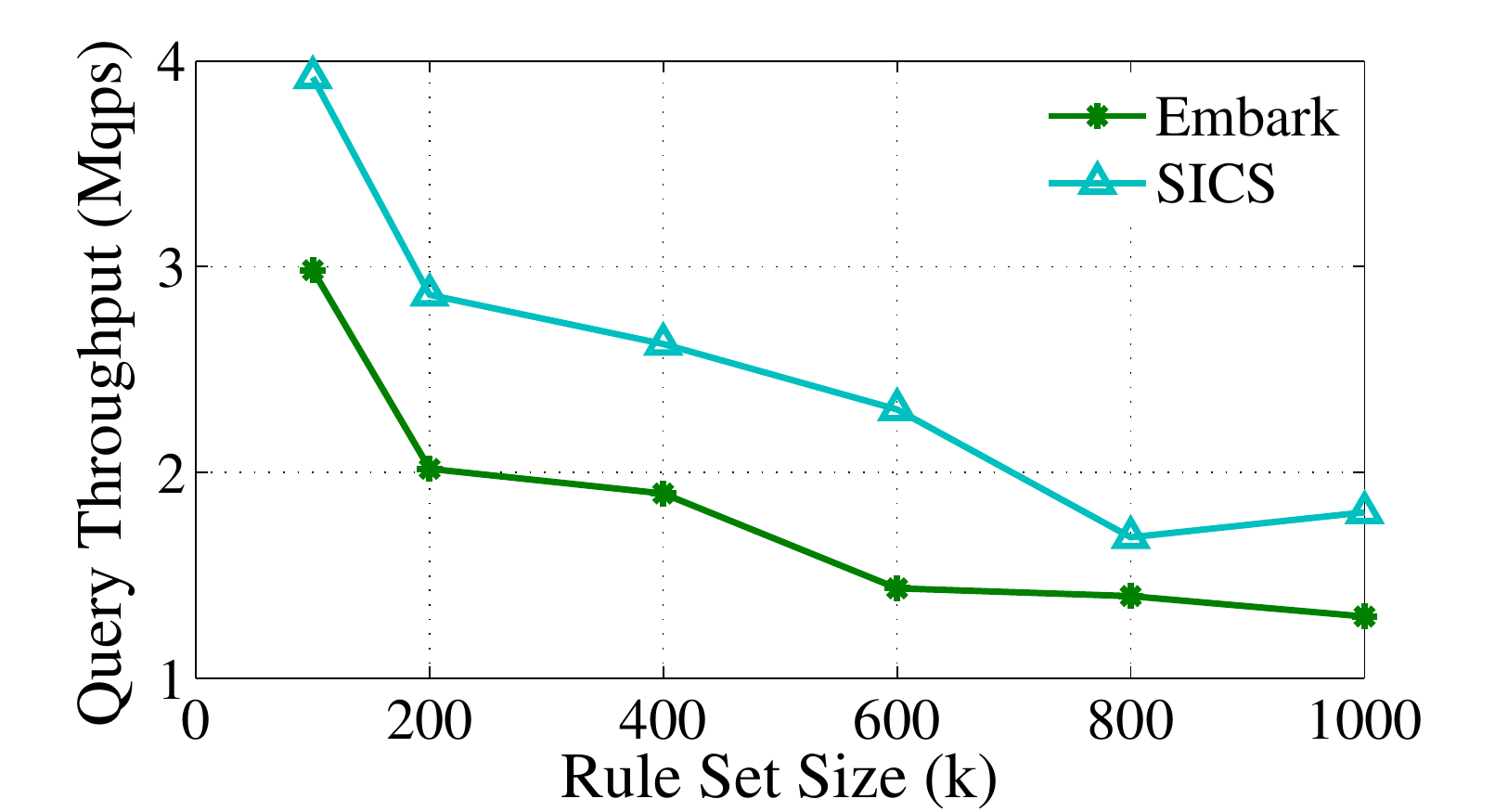}
\caption{Query troughput as \# of rules increases.}
\label{fig:throughput}
\vspace{-1ex}
\end{figure}

\textbf{Query throughput.}
In this set of experiments, we measure the query throughput of the gateway in SICS, in a number of queries per second (qps). Packets used for queries in the experiments are generated randomly with respect to equivalence classes. Results for each ruleset are shown in Fig.~\ref{fig:throughput}.
Since the PredixMatch in Embark can support only one header field. Incoming packets need to query for each packet header field sequentially.
From the figure, we see that the gateway in SICS can achieve 3.92 Mqps for the ruleset with 100K rules. For the largest ruleset with 1000K rules, the query throughput is 1.5 Mqps. For all rulesets, query throughput of the gateway in SICS is higher than that in Embark by approximately 20\%.

\begin{figure}[t]
\centering
\includegraphics[width=0.8\linewidth]{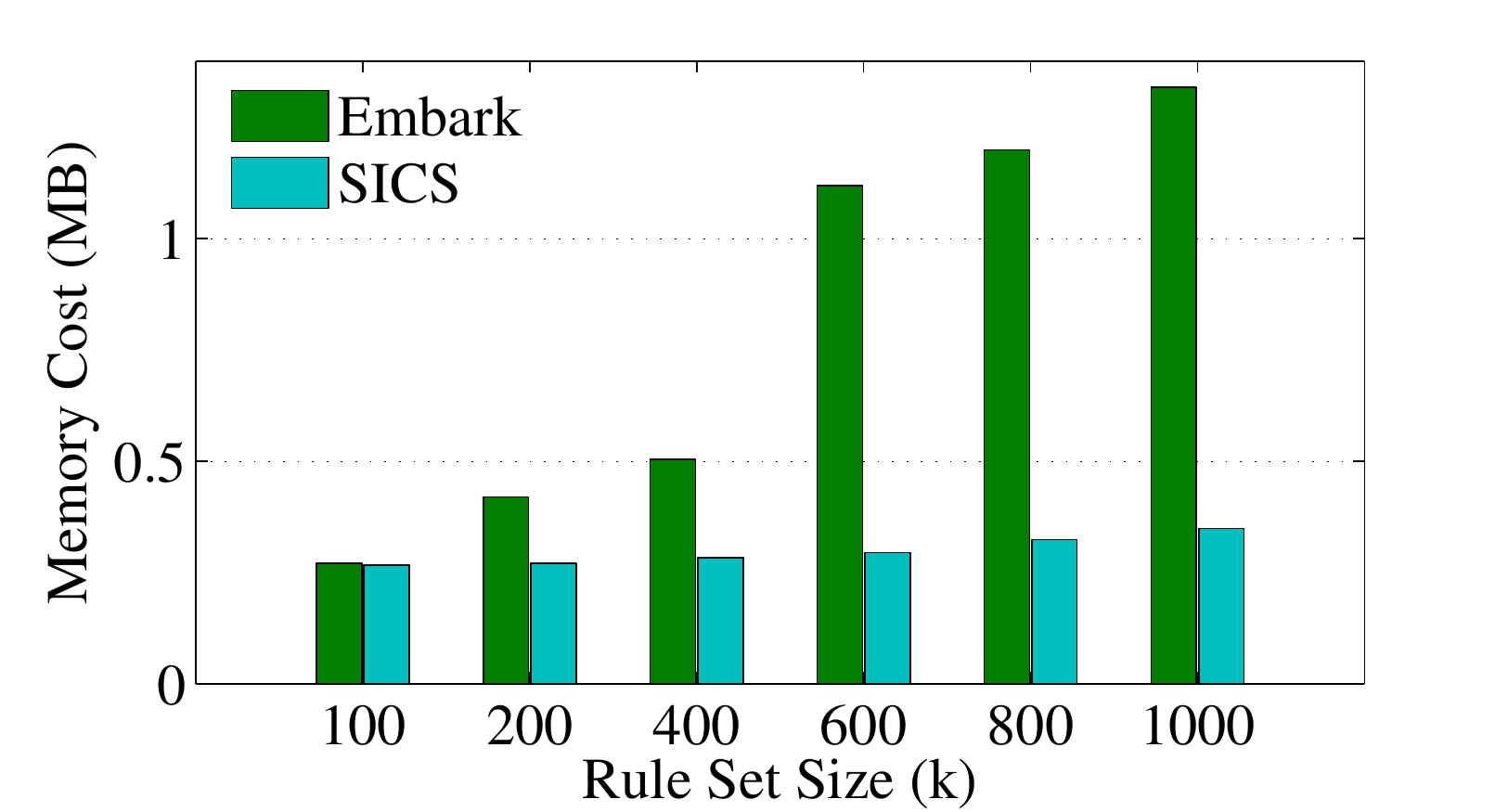}
\caption{Memory usage of the gateway.}
\label{fig:memory}
\vspace{-1ex}
\end{figure}

\textbf{Memory usage.}
Fig.~\ref{fig:memory} shows the memory usage of the gateway.
The gateway in SICS does not store rules. Instead, we only store predicates calculated by the rule composition module. Predicates are represented as BDDs in our implementation.
For each predicate, we maintain a representation list recording a subset of equivalence classes and their corresponding labels whose disjunction is equal to the predicate.
Each equivalence class is represented as a set of pointers to predicates which contain the equivalence class.
For in Embark, we only calculate the memory cost of the data structure for PrefixMatch.
As shown in Fig.~\ref{fig:memory}, the gateway in SICS use less memory than Embark for all rule sets.
The memory usage for 100K rules is 0.267 MB and it increases to 0.349 MB when the ruleset becomes 10 times larger.
The gateway costs very small memory and can be stored in the cache.

\subsection{Bandwidth Overheads}
\label{sec:bandwidth}
We evaluate the bandwidth overhead due to SICS encryption.
In Embark, the PrefixMatch increases the amount of data sent to the cloud by 20-bytes per packet (IPv4 to IPv6 conversion).
Compared with Embark, the encryption scheme in SICS does not incur extra bandwidth overheads on packets.
In our implementation, we use 16 bits to encode labels. A 16-bits long label can support up to 65536 equivalence classes (network-wide behaviors) which are enough for a middle-sized enterprise network. As reported in \cite{APVerifier}, Stanford backbone network which contains 757170 forwarding rules has 494 network-wide behaviors.
For middleboxes which modify packet headers, we use another 16 bits as identifiers in a header fields pool to identify rewritten header fields.
Summing up above, the total number of bits we use in our encryption scheme is 32 bits which can be placed in the options field of IPv4 protocol header. So for each packet, no extra bandwidth overhead is added.

\subsection{In-cloud Middleboxes}
\label{sec:middlebox}

\begin{figure}[t]
\centering
\includegraphics[width=0.8\linewidth]{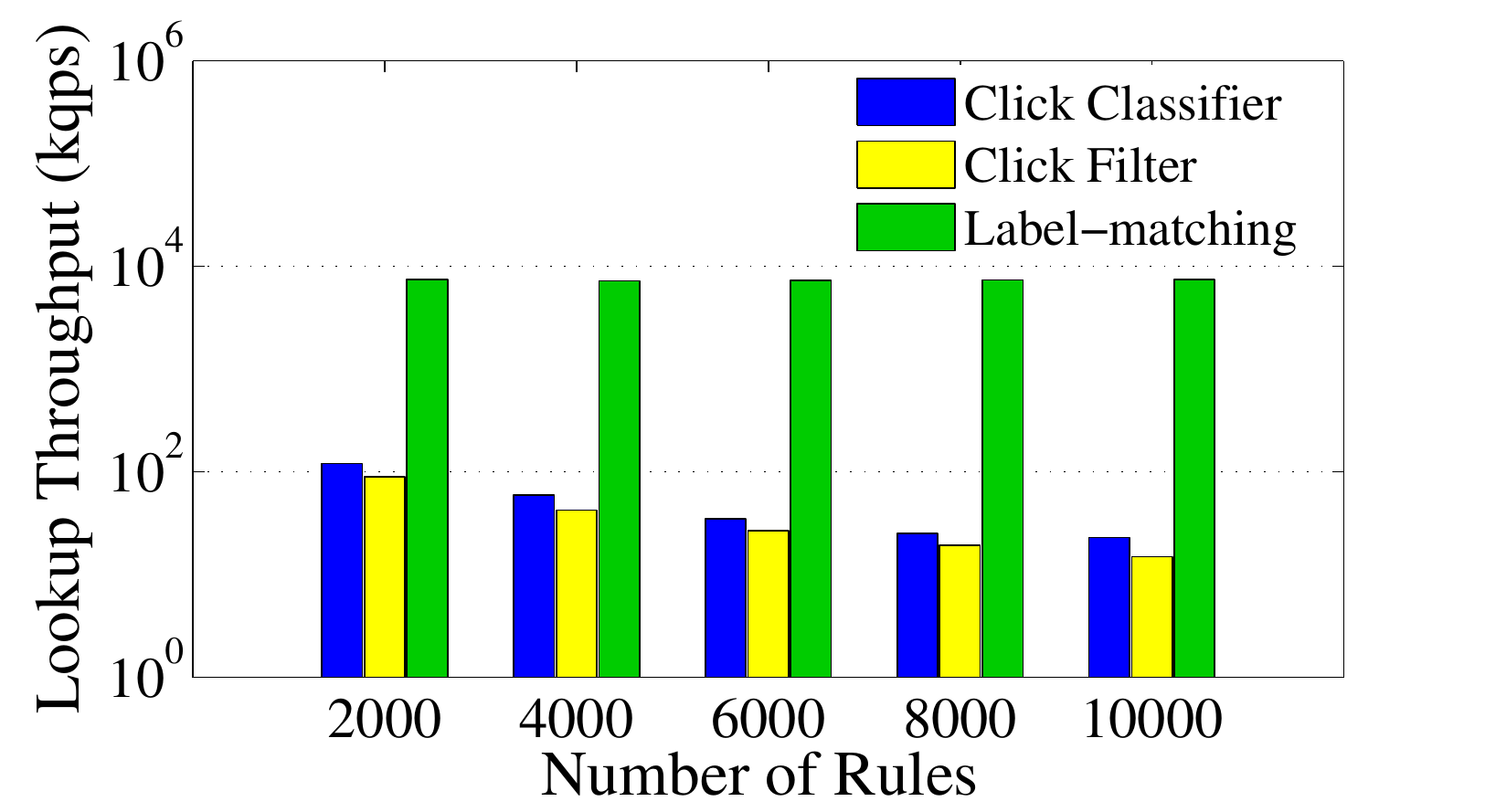}
\caption{Lookup throughput of Middleboxes.}
\label{fig:middlebox}
\vspace{-1ex}
\end{figure}
In this set of experiments, we evaluate the performance of our label-matching based middleboxes in the cloud.
We develop middleboxes using Click modular router.
The label-matching Click element is implemented as a (2,4)-Cuckoo hash table which uses 64 KB memory. Keys are 16-bits long labels. Their corresponding values indicate behaviors of packets with these labels.

\textbf{Query throughput of label-matching.}
We compare the query throughput of label-matching with two built-in elements in Click: Click Classifier and Click Filter. Click Classifier classifies packets based on patterns of packet headers. Click Filter denies or permits packets based on five tuples of packet headers.
Fig.~\ref{fig:middlebox} shows the experimental results with the number of rules on the middlebox increasing from 2000 to 10000.
The y axis is the lookup throughput in thousand of queries per second (log scale). From the figure, we see that label-matching can achieve about 8M queries per second, which is larger than Click Filter and Click Classifier by about two orders of magnitude.

\begin{figure}[t]
\centering
\includegraphics[width=0.8\linewidth]{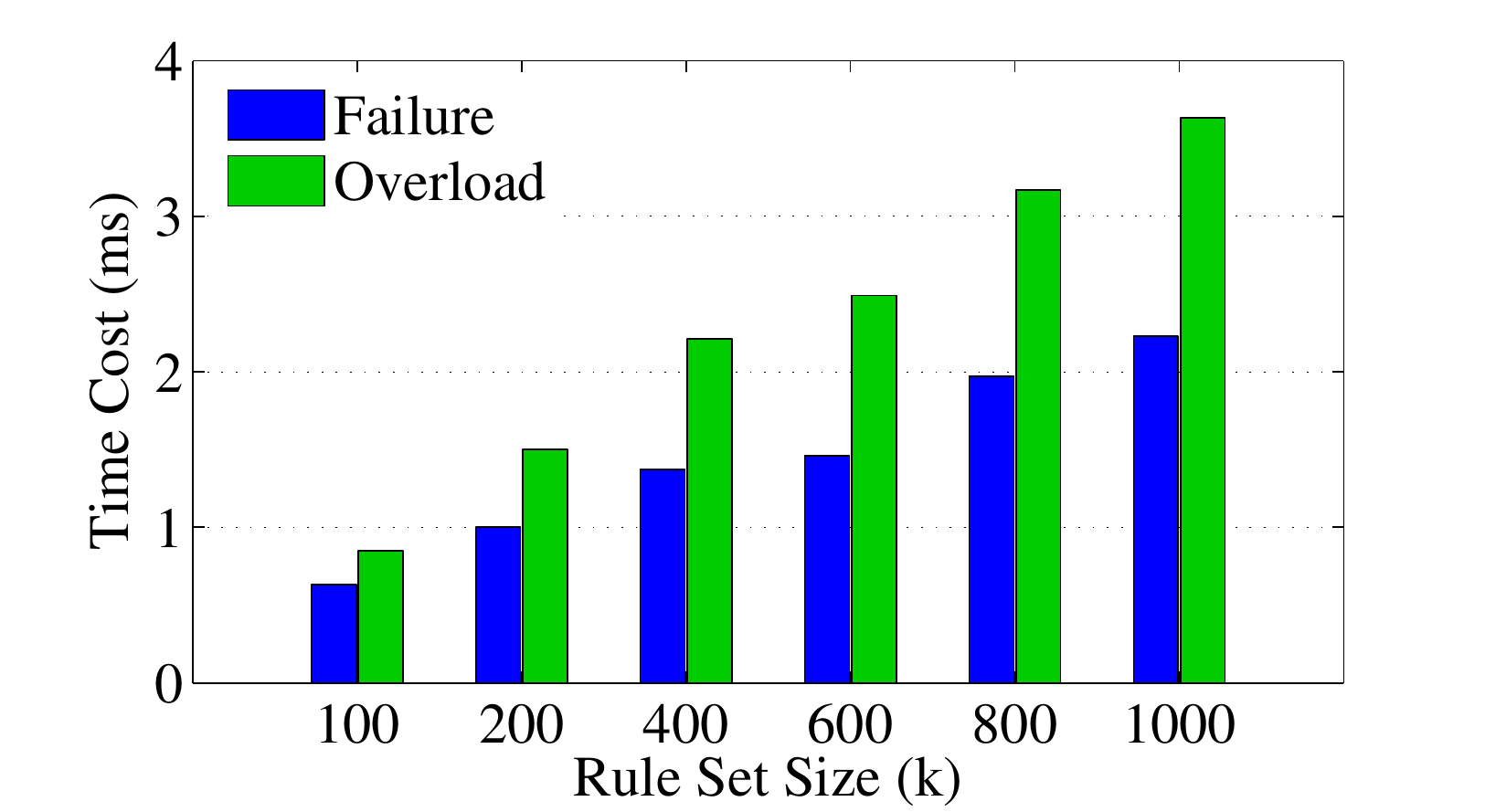}
\caption{Response time in the case of a middlebox failure and traffic overload. }
\label{fig:overload}
\vspace{-2ex}
\end{figure}

\textbf{Reacting to middlebox failure and overload.}
We consider two dynamic scenarios: 1) a middlebox fails and 2) traffic overload at a middlebox. We measure the reacting time of SICS for each scenario. Results are shown in Fig.~\ref{fig:overload}. When a middlebox fails, we need to migrate the state of the failure middlebox to a new instance and reconfigure the network to steer packets with certain labels to the new instance.
For traffic overload at a middlebox, besides middlebox state migration, we need to add new predicates to reroute a portion of traffic on the current middlebox to another middlebox. This further requires updating the packet classifier and representation lists at the gateway.
From the Fig.~\ref{fig:overload}, we see that the overall time to react to middlebox failure and traffic overload is low (several milliseconds) and it costs a little more time to deal with middlebox overload than failures.

\section{Conclusion}
\label{sec:conclusion}

SICS is a middlebox outsourcing system that protects the private information of packet headers and middlebox rules. Compared to existing methods, SICS has several unique advantages including stronger security guarantee, high-throughput processing at both enterprise and cloud sides, and support of fast update. SICS assigns each packet a label identifying its matching behaviors in a service chain and all middlebox processing in the cloud will be based on  labels. We use prototype implementation and evaluation  on Amazon VPC and local computers to show the feasibility, high performance, and efficiency of SICS.

\newpage

{\small
\bibliographystyle{abbrv}
\bibliography{bibfile}}




\end{document}